\documentclass[12pt,preprint]{aastex}
\usepackage{color}

\begin{document}

\title{A study on the sharp knee and fine structures of cosmic ray spectra}

\author{Bo Wang\altaffilmark{1}, Qiang Yuan\altaffilmark{1},
Chao Fan\altaffilmark{2,1}, Jian-Li Zhang\altaffilmark{1}, Hong-Bo
Hu\altaffilmark{1}, Xiao-Jun Bi\altaffilmark{1}}

\affil{$^1$Key Laboratory of Particle Astrophysics, Institute of High
Energy Physics, Chinese Academy of Sciences, Beijing 100049, P.R.China\\
$^2$Department of Physics, Shandong University, Jinan 250100, P. R. China}

\begin{abstract}
The paper investigates the overall and detailed features of cosmic
ray (CR) spectra in the knee region using the scenario of
nuclei-photon interactions around the acceleration sources. Young
supernova remnants can be the physical realities of such kind of CR
acceleration sites. The results show that the model can well explain
the following problems simultaneously with one set of source
parameters: the knee of CR spectra and the sharpness of the knee,
the detailed irregular structures of CR spectra, the so-called
``component B'' of Galactic CRs, and the electron/positron excesses
reported by recent observations. The coherent explanation serves as
evidence that at least a portion of CRs might be accelerated at the
sources similar to young supernova remnants, and one set of source
parameters indicates that this portion mainly comes from standard
sources or from a single source.
\end{abstract}

\keywords{cosmic rays -- knee -- fine structures -- e$^+$e$^-$
excesses--``component B''--single source}

\maketitle

\section{introduction}
Since the discovery by \cite{KK1958}, the ``knee'' of cosmic ray
spectra has been one fundamental problem of CR physics for half a
century. Many theoretical works try to explain this interesting and
important phenomenon. The most popular explanation attributes the
knee to the inefficient acceleration of the Galactic CRs by the
accelerators above PeV energies
\citep[e.g.,][]{2002PhRvD..66h3004K,2003A&A...409..799S,2001JPhG...27.1005E}.
Alternative possibilities include the leakage of CRs when
propagating in the Galaxy
\citep{1993A&A...268..726P,2004IJMPA..19.1133R,
2001NuPhS..97..267L}, interactions between CRs and the background
light \citep{1993APh.....1..229K,2002APh....17...23C} or neutrinos
\citep{2003APh....19..379W} before arriving at the earth, or exotic
interaction of CRs in the atmosphere where undetectable particles
are produced and missing the detection \citep{2000PAN....63.1799N,
2001ICRC....5.1760K}.

From the experimental aspects, the measurements of the CR spectra
around the knee region become increacingly precise, which can even
reveal some fine structures of the knee. After a long term
operation, the Tibet Air Shower array reported a very good
measurement of the knee spectra with unprecedented high statistics
and low systematics \citep{2008ApJ...678.1165A}. Especially
interesting signature of the Tibet result is that the CR spectra
show a very sharp break of the spectrum index around 4 PeV. At
almost the same time several experiments have reported their new
measurements with the similar behavior, such as
KASKADE\citep{2009APh....31...86A},
ARAGATS-GAMMA\citep{2008JPhG...35k5201G},
Yakutsk\citep{2009NJPh...11f5008I}, and
MAKET-ANI\citep{2007APh....28...58C}. Such a sharp knee challenges
the traditional interpretations of the knee
\citep{Shibata2009,2009arXiv0906.3949E,2009arXiv0911.3034H}. It is
shown that if adopting an exponential-like cutoff of each component
of CR species with low He flux, it will be very difficult to
reproduce the sharp knee data \citep{Shibata2009}. It is suggested
by \cite{2009arXiv0911.3034H} that a double power-law may well fit
the observational data with high He flux, which indicates that He
may be the main component around the knee. Furthermore Hillas
suggested that there should be another Galactic component,
``component B'', to explain the CR spectra above $10$ PeV
\citep{2005JPhG...31R..95H}.

Besides the sharp transition of the knee, \cite{2009arXiv0906.3949E}
carefully analyzed the CR spectra of individual experiment. By
renormalizing the energy with respect to the break point(measured
knee energy)of individual experiment, the problem related to the
uncertainty of the absolute energy scale can be avoided, so the
deviations of the observed spectra from the fitted spectra can be
combined for all experiments. The result clearly shows the
peculiarities at the positions expected for CNO group and Fe group
if the knee is corresponding to the position of He. Interestingly
the energies of these fine bumps are proportional to the mass number
of the several major nuclei species: proton, He, CNO and Fe. The
sharp knee and the irregularities of CR spectra are regarded as
evidence for the single source origin of CRs
\citep{2009arXiv0906.3949E}.

Another important development in CR physics is the new discovery of
electron/positron excesses by several experiments
\citep{2009Natur.458..607A,2008Natur.456..362C,2008PhRvL.101z1104A,
2009A&A...508..561A,2009PhRvL.102r1101A}. To explain the positron
fraction and electron spectrum excesses simultaneously one may need
to introduce some exotic sources of e$^+$e$^-$ pairs
\citep{2009PhRvD..79b1302S}. \cite{2009ApJ...700L.170H} (hereafter
Paper I) proposed a model resorting to e$^+$e$^-$ pair through
interactions of CR nuclei and ambient photons around the
acceleration sources, which can explain the knee of the CR spectra
and the electron/positron excesses at the same time. Based on that
model we further study the detailed structures of the CR spectra in
this work, intending to reproduce the sharp knee and fine structures
mentioned above. In our interaction model the threshold energy of
different chemical compositions is $A$-dependent, which will result
in an $A$-dependent knee of each composition. This feature is
consistent with the property of the fine structures found in
experimental data \citep{2009arXiv0906.3949E}. In addition, the
interaction will cause a pile-up of the particles below the
threshold. We expect this effect can contribute to the sharp knee
and irregular bumps of the CR spectra.

This paper is organized as follows. In Sec. 2 we will first go over
the model describing CR-photon interactions. In Sec. 3 we present
the calculated results and comparisons with observational data of
the sharp knee and irregular structures of CR spectra. Finally Sec.
4 is the conclusion.

\section{Interactions between CRs and ambient photons}
The model to explain the knee and electron/positron excesses using
nuclei-photon interactions around the acceleration sources is
proposed in Paper I. Here we readdress the basic physical picture
and give some technical details.

\begin{figure}[!htb]
\begin{center}
\includegraphics[width=8cm]{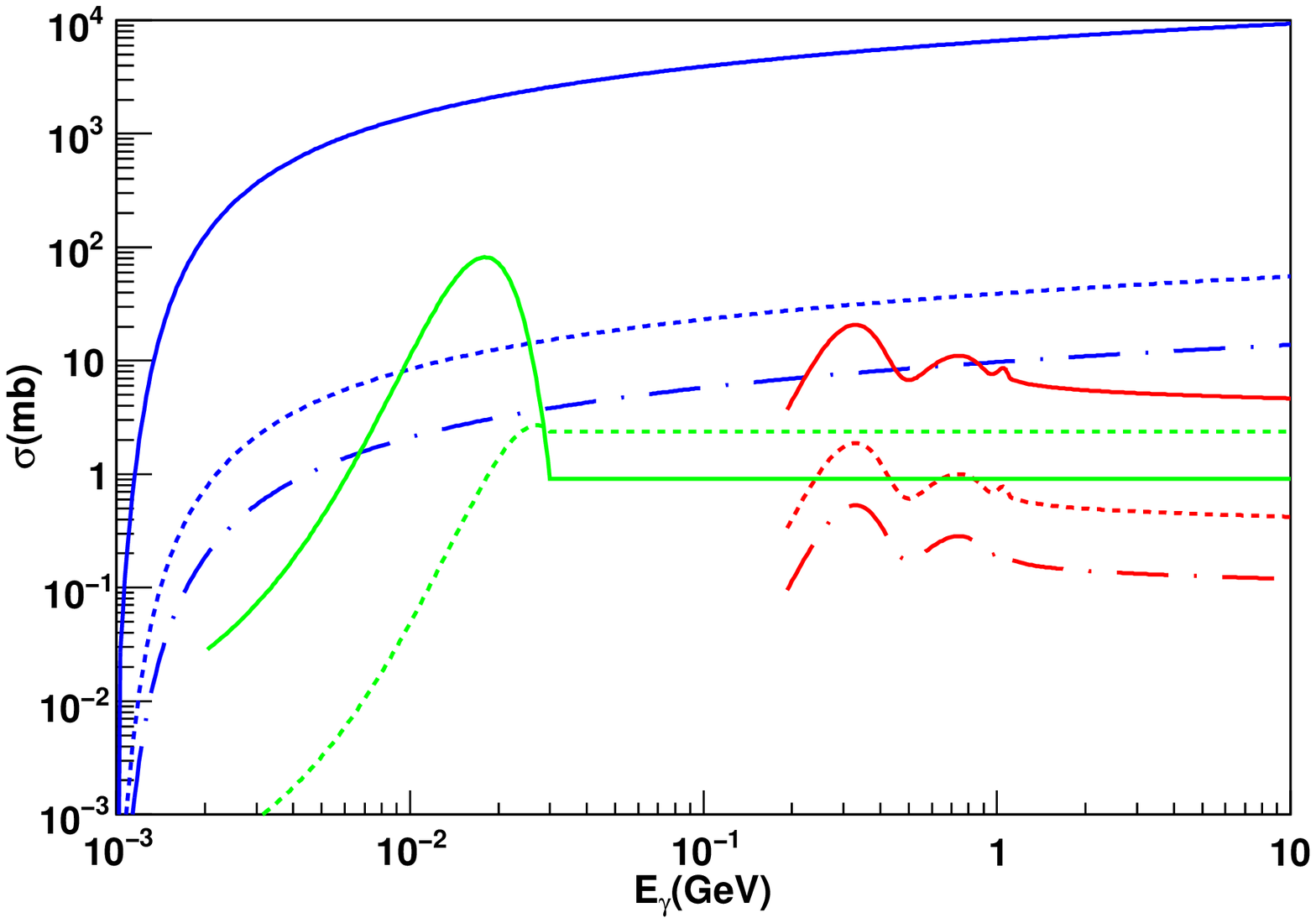}
\includegraphics[width=8cm]{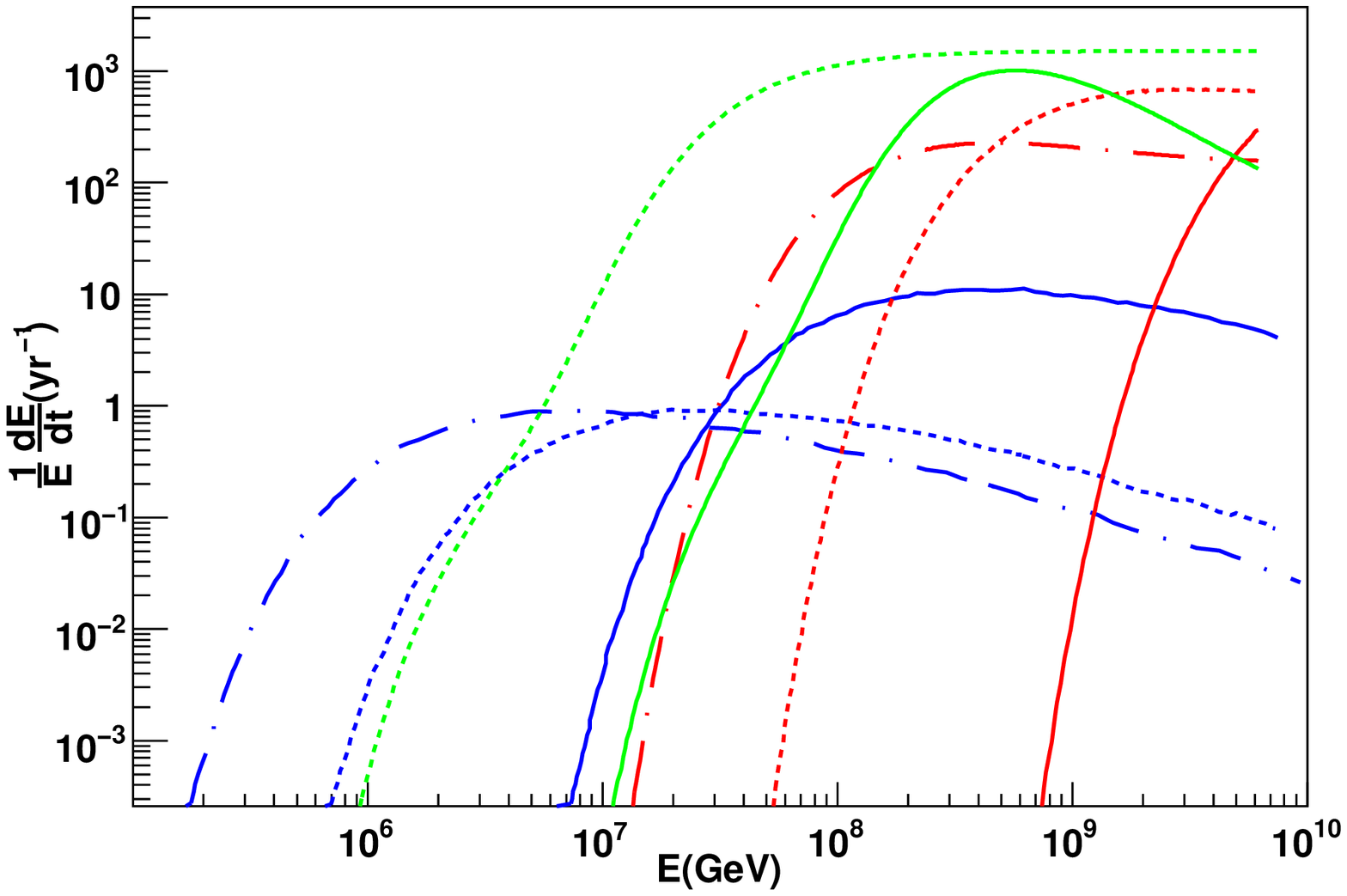}
\caption{{\it Left:} cross
 sections as functions of photon energy
in the nuclei rest system for proton (dash-dotted), He (dashed) and
Fe (solid) respectively. For each kind of nuclei, the cross sections
for the three interaction processes: pair production (blue),
photodisintegration (green) and pion production (red) are shown.
     {\it Right:} the average relative energy loss rates
$\frac{1}{E}\frac{{\rm d}E}{{\rm d}t}$ for proton, He and Fe due to
pair production, photodisintegration and pion production
interactions in a $5000$ K blackbody radiation field with respect to
the nuclei energy in laboratory system. The labels of lines are same
as in the left panel. } \label{cross_sec}
\end{center}
\end{figure}

There are three kinds of interactions between CRs and photons: pair
production, photodisintegration and photo-pion productions processes
when very high energy CR nuclei interact with background photons.
The cross sections for pair production and photodisintegration are
given in \cite{1970PhRvD...1.1596B} and \cite{1976ApJ...205..638P}
respectively. The pion production cross section for proton is
adopted from \cite{2008PhLB..667....1P}, and we employ an $A^{0.91}$
dependence for other nuclei with atomic number $A$
\citep{1985PhRvD..32.1244S}. The cross sections as functions of
photon energy in the nuclei rest system for proton, He and Fe, which
are the dominant compositions for CRs around the ``knee'' region
\citep{2003APh....19..193H}, are shown in the {\it left panel} of
Fig. \ref{cross_sec}. Note that for proton there is no
photodisintegration interaction. The pair production cross section
is proportional to $Z^2$ of the nuclei, so for heavy nuclei like Fe
the pair production cross section is extremely large
\citep{Blumenthal:1970nn}.

To see clearly the effects of the three interaction channels, we
need to know the energy loss of the nuclei due to each of the
interactions. The energy loss rate can be written as

\begin{equation}
\frac{{\rm d}E}{{\rm d}t}=\frac{\kappa E}{\tau(E)}=\kappa E\cdot
{\int {\rm d}\cos\theta\frac{1-\cos\theta}{2} \int {\rm d}\epsilon
\,n(\epsilon)\,\sigma(E,\epsilon,\cos\theta)c},
\end{equation}

where $\tau(E)$ is the average interaction time of the nuclei in a
radiation field with number density $n(\epsilon)$, $\theta$ is the
angle between photon and nuclei momenta, $c$ is speed of light, and
$\kappa$ is the average fraction of energy loss in one collision,
i.e. the inelasticity. We plot in the {\it right panel} of Fig.
\ref{cross_sec} the average relative energy loss rates with respect
to the nuclei energy in laboratory system due to each of the
interactions for proton, Helium and Iron respectively. The
background radiation field in the calculation is specfied to be
blackbody field with $T=5000$ K. The average inelasticity $\kappa$
for pair production is calculated using a Monte-Carlo (MC) study to
average the output e$^+$e$^-$ pair energies for many realizations.
For photodisintegration process the inelasticity is simply adopted
as $\kappa=i/A$ with $A$ the total number of nucleus and $i$ the
number of nucleons kicked out in one interaction. It is shown that
$i=1$ is dominant for photon energy below $30$ MeV in the nuclei
rest system \citep{1976ApJ...205..638P}. For pion production the
average inelasticity is adopted as $\kappa=
\frac{1}{2}\left(1+\frac{m_{\pi}^2-m_A^2}{s}\right)$ with
$m_{\pi},\, m_A$ the masses of pion meson and nuclei, and $s$ the
center-of-moment system energy \cite{1968PhDT.........3S}. We can
see from Fig. \ref{cross_sec} that for energies lower than several
PeV, the pair production dominates the energy losses for proton and
Helium. The photodisintegration process of Helium become important
above $\sim 4$ PeV. Pion production is only important for energies
higher than several tens PeV.

We use MC method to simulate the interactions between CR nuclei and
ambient photons. The acceleration processes and interactions are
decoupled for simplicity. The CR nuclei are assumed to inject into
the radiation field with power-law spectra and proper relative
abundance according to the measurements
\citep{2003APh....19..193H,2008JPhCS.120f2023B}. Note here we adopt
a correction of the measured spectra to the source spectra, taking
into account the propagation effect (see below). After interactions
the CRs together with the e$^+$e$^-$ products enter the interstellar
environment and propagate diffusively in the Galaxy. For the
propagation of CR nuclei we simply use the leaky-box model with
escape time $\tau_{\rm esc}(R)\approx2\times10^8\left(\frac{R}{1{\rm
GV}}\right)^{-0.6}$ yr \citep{2009A&A...497..991P}. The propagation
of e$^+$e$^-$ is a bit complicated since the dominant effect is
energy loss due to synchrotron radiation and inverse-Compton
scattering, instead of diffusion (or to say escape). We adopt
GALPROP code to calculate the propagation of electrons and
positrons.

The radiation field around the acceleration source is assumed to be
of blackbody shape, with a temperature several to ten thousand
Kelvin. It might be true for the young supernova shortly after its
explosion. The intensity of the radiation field does not have to be
as intense as the blackbody radiation. We keep the density of
photons as a free parameter. The photon density multiplied with the
interaction time gives the effective interaction probability.

As we have stated in Paper I, the photodisintegration of heavy
nuclei, especially He, will overproduce protons through secondary
production of protons. Therefore we do not expect the
photodisintegration to play a significant role in the interaction
process. We find that an evolving picture with decaying and
asymptotically cooling of the radiation field can naturally explain
this requirement. Schematically speaking the photodisintegration and
pion production have higher threshold, which requires longer time
for CR nuclei to be accelerated to exceed the threshold. But as time
goes on the temperature of the radiation field may become lower and
lower, and the interactions of photodisintegration and pion
production need even higher energies of CRs. Therefore the
photodisintegration and pion production can be effectively
suppressed, and only the pair production takes effect.

\section{Results}

he calculated energy spectra of individual composition, including
proton, He, CNO and Fe, together with the observational data are
shown in Fig.2. The main parameters used in the MC calculation,
including the relative abundances and spectral indices at the
source, temperatures of radiation field, photon column density, and
equivalent time for blackbody field, are compiled in Table 1. The
relative abundances are adopted to reproduce the locally observed
fluxes of each species. As same as in Paper I, only one set of
parameters is enough to describe all spectra well. A few model
parameters are slightly adjusted in order to better agree with the
fine structures of CR spectra(see detailed explanation below).
However, the basic features as shown in Paper I are kept unchanged.
The power-law index and normalization of each chemical composition
are the same as that used in Paper I, except for CNO group we use a
harder injection spectrum by $0.06$ and a lower flux normalization
at 1 TeV by $10\%$ according to \cite{2008JPhCS.120f2023B}, which
will make the flux slightly go up around the knee region. The
temperatures of the radiation field for proton and He, which are the
main compositions of CRs, reamin the same as in Paper I. However, we
change the temperature for Fe from $\sim 2000$ K in Paper I to $\sim
3500$ K in this work. In Paper I we set a lower temperature for Fe
with the purpose that the second knee around $300-400$ PeV might be
reproduced using the drop of Fe spectrum. If we relax this
requirement we have shown that a wide range of the temperature for
Fe can be consistent with the current data (see the systematic check
done in Paper I). As we will see below, a higher temperature for Fe
will be better to reproduce the fine structures of the total CR
spectrum. For other nuclei which do not play a significant role, we
use the temperature field $5500$ K for $3\leq Z \leq 25$ and $3500$
K for $Z\geq 26$ respectively. It is shown that the model
expectations are consistent with the observational data. However,
since the current measurements of individual CR composition are
actually not very good, the constraints on the model parameters are
weak.

\begin{figure}[!htb]
\centering
\includegraphics[angle=-0,width=0.8\columnwidth]{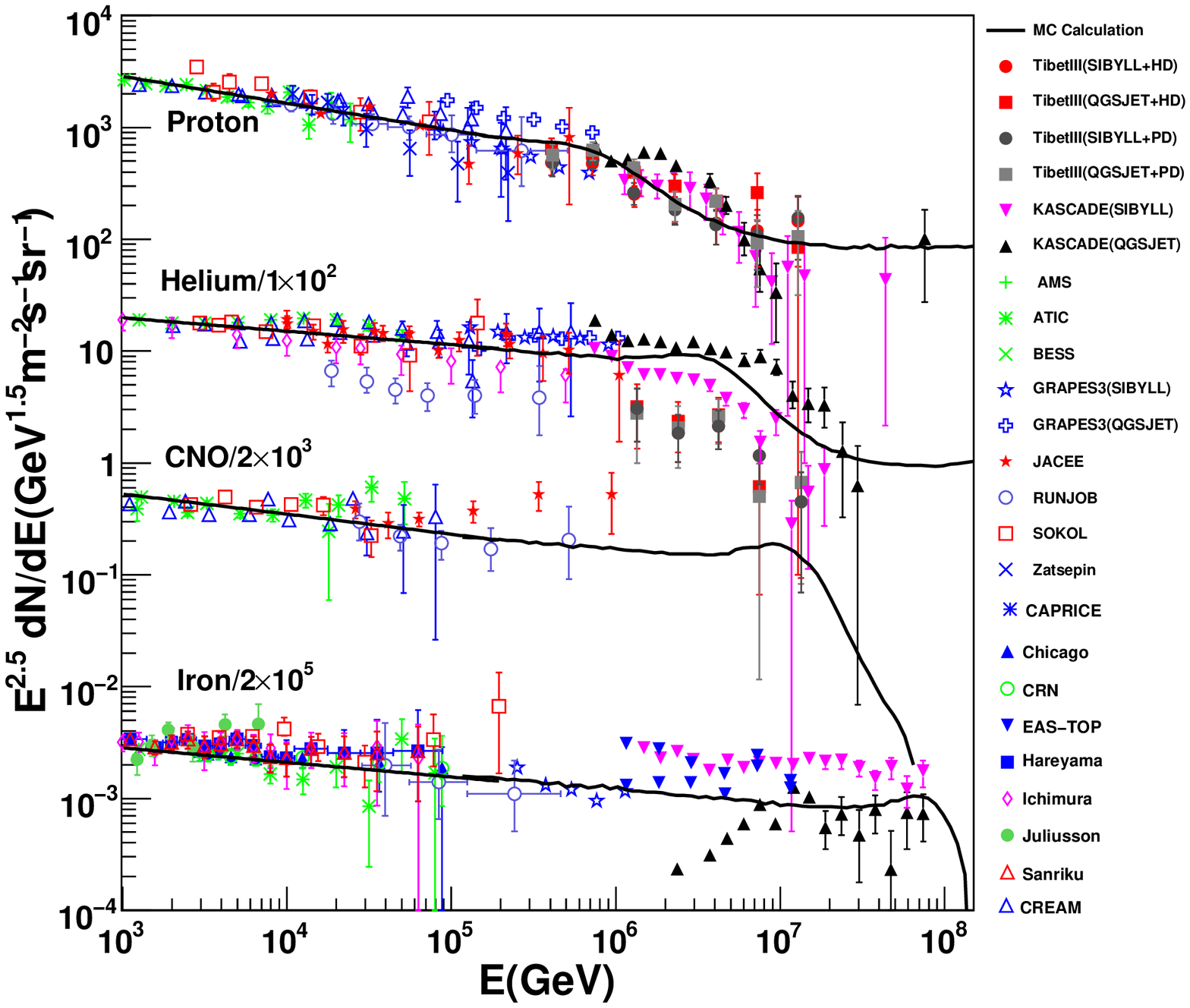}
\caption{\small Energy spectra for proton,
 He, CNO and Fe. The solid
lines are the MC calculated results. Observational data are
Tibet-III \citep{2006PhLB..632...58T}, KASCADE
\citep{2005APh....24....1A, 2009APh....31...86A}, AMS
\citep{2000PhLB..490...27A,2002PhR...366..331A}, ATIC-2
\citep{2009BRASP..73..564P}, BESS \citep{2004PhLB..594...35H},
GRAPES-3 \citep{2008ICRC....5.1121G}, JACEE
\citep{1998ApJ...502..278A}, RUNJOB \citep{2001APh....16...13R},
SOKOL \citep{1993ICRC....2...17I}, Zatsepin
\citep{1993ICRC....2...13Z}, CAPRICE \citep{2003APh....19..583B},
Chicago \citep{1991ApJ...374..356M,1995ICRC....2..652S}, CRN
\citep{1991ApJ...374..356M}, EAS-TOP, Hareyama
\citep{1999ICRC....3..105H}, Ichimura \citep{1993PhRvD..48.1949I},
Juliusson \citep{1974ApJ...191..331J}, Sanriku
\citep{1997APh.....6..155K}, CREAM
\citep{2008ICRC....2...55Y,2009ApJ...707..593A};} \label{Individual}
\end{figure}

\begin{table}
\begin{center}\caption{Parameter settings in the Monte-Carlo
calculation.}
\begin{tabular}{lccccc}
\hline \hline  & relative abundance & $\gamma_{_{Z}}$ & $T_{\rm ph}$
& $\langle nc\tau\rangle$ & $\hat{\tau}$
\vspace{-0mm} \\
  & $10^5$ GeV$-10^9$ GeV  &  & (K)& ($10^{29}$cm$^{-2}$) & (yr) \\
\hline
 Proton  & $1.00$ & $2.14$ & $1.0\times 10^4$ & $8.1$ & $0.04$   \\
  Helium  & $0.66$ & $2.02$ & $7.0\times 10^3$ & $12.9$ & $0.19$  \\
  CNO     & $0.30$ & $2.02$ & $5.5\times 10^3$ & $7.0$ & $0.21$   \\
  Iron    & $0.23$ & $2.03$ & $3.5\times 10^3$ & $2.0$ & $0.23$   \\
  \hline
  \hline
\end{tabular}
\label{table1}
\end{center}
\end{table}

The case for the all-particle spectra is much better than the
individual composition. The comparison of the all-particle spectra
between our theoretical prediction and the data is shown in Fig.3 We
can see the knee forms due to the break of He spectrum around $\sim
4$ PeV. This condition would require the temperature of radiation
field which He experiences is about $7000$ K. The pile-up of He
particles around the threshold point helps to better produce the
sharp break of the knee spectrum. To investigate the break behavior
more quantitatively, we adopt the following double power-law
function to fit the CR spectrum below and above the knee
\citep{2009arXiv0906.3949E}

\begin{equation}
I(E)=AE^{\gamma}\left[1+\left(\frac{E}{E^{k}}\right)^{\delta}\right]
^{-\frac{\Delta\gamma}{\delta}}, \label{DoublePowLaw}
\end{equation}
where $\gamma$ is the spectrum index below the knee, which changes
by $\Delta \gamma$ above the knee, $\delta$ is the sharpness
parameter which describes the smoothness of the transition. The
sharpness $S$ (the second-order derivative of spectrum at the energy
break point E$_{k}$) is defined as \citep{2009arXiv0906.3949E}

\begin{equation}
S=\delta\Delta \gamma \frac{\ln10}{4}.
\end{equation}
Through fitting the theoretical calculated spectrum we can obtain
the sharpness $S = 2.0$, while the sharpness of experimental data of
Tibet air shower array is about $2.4\pm 0.8$, which shows good
consistence with the theoretical one.

\begin{figure}[!htb]
\centering
\includegraphics[angle=-0,width=0.8\columnwidth]{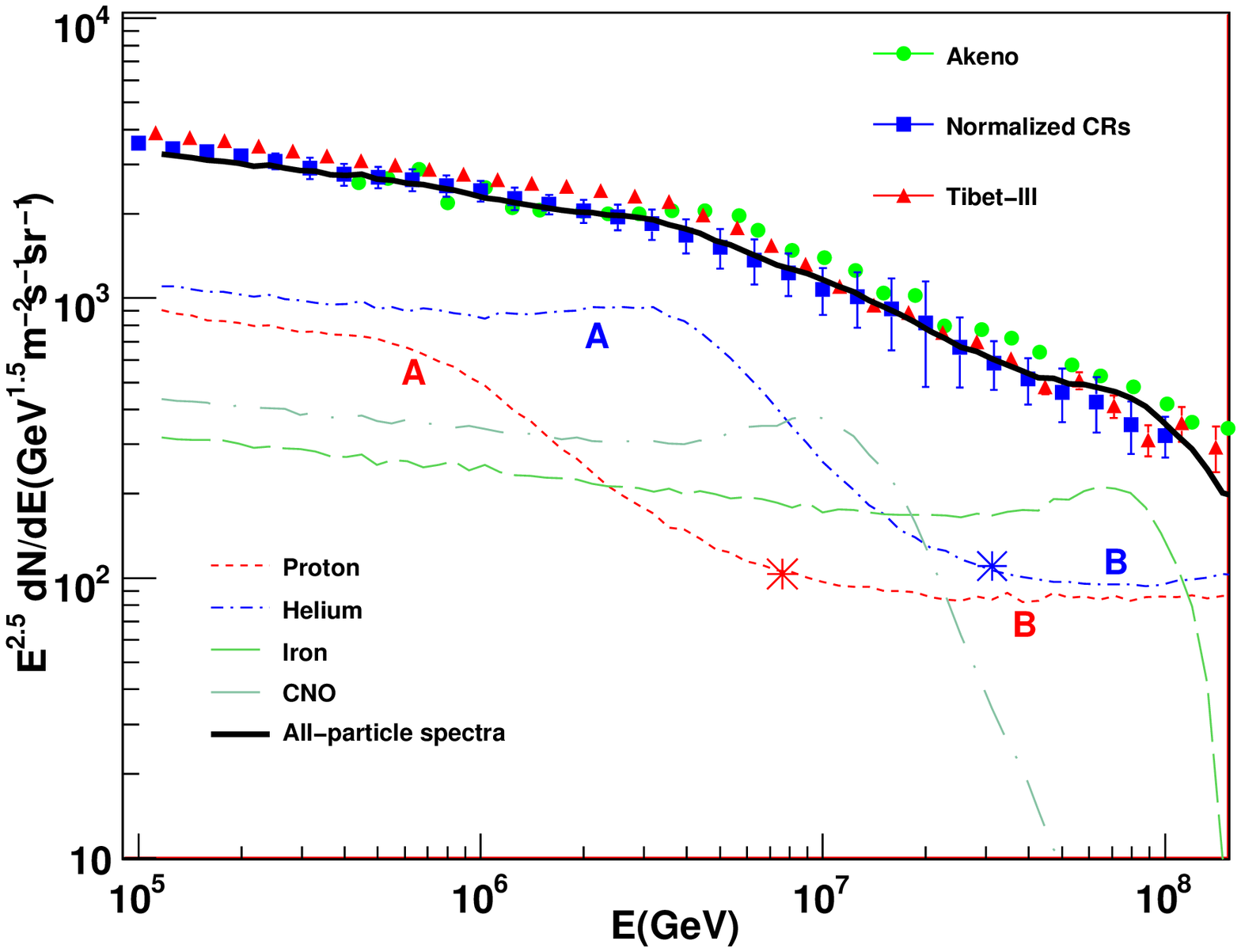}
\caption{\small The calculated all-particle spectrum. Also shown are
proton, He, CNO and Fe components. The observational data:
Tibet-III\citep{2008ApJ...678.1165A}, Akeno
\citep{1984JPhG...10.1295N}. The {\it Normalized} data are derived
by combing all data with a rescale based on the extrapolation of the
direct measurements\citep{2003APh....19..193H}. Label ``B''
indicates the high energy tails of the proton and He spectra above
marker ``*'' can be regarded as the ``component B'' of the Galactic
CRs as suggested by \cite{2005JPhG...31R..95H}. }
\label{AllParticleKnee}
\end{figure}

We can also note that there are high energy tails of the proton and
He spectra due to the decrease of the pair production energy loss
rate (Fig. \ref{cross_sec}). This behavior is very similar to the
ankle of the ultra high energy CRs generated by e$^+$e$^-$ pair
production interactions with cosmic microwave background photons
\citep{2006PhRvL..97w1101B}. Coincidentally, it may be a natural
explanation of the ``component B'' of the Galactic CRs as suggested
by \cite{2005JPhG...31R..95H} when studying the overall property of
CR spectrum. However for the heavy nuclei, there are no obvious high
energy tails due to large cross sections and the constant
temperature of individual component, as shown in \ref{cross_sec}.

\begin{figure}[!htb]
\centering
\includegraphics[width=0.9\columnwidth]{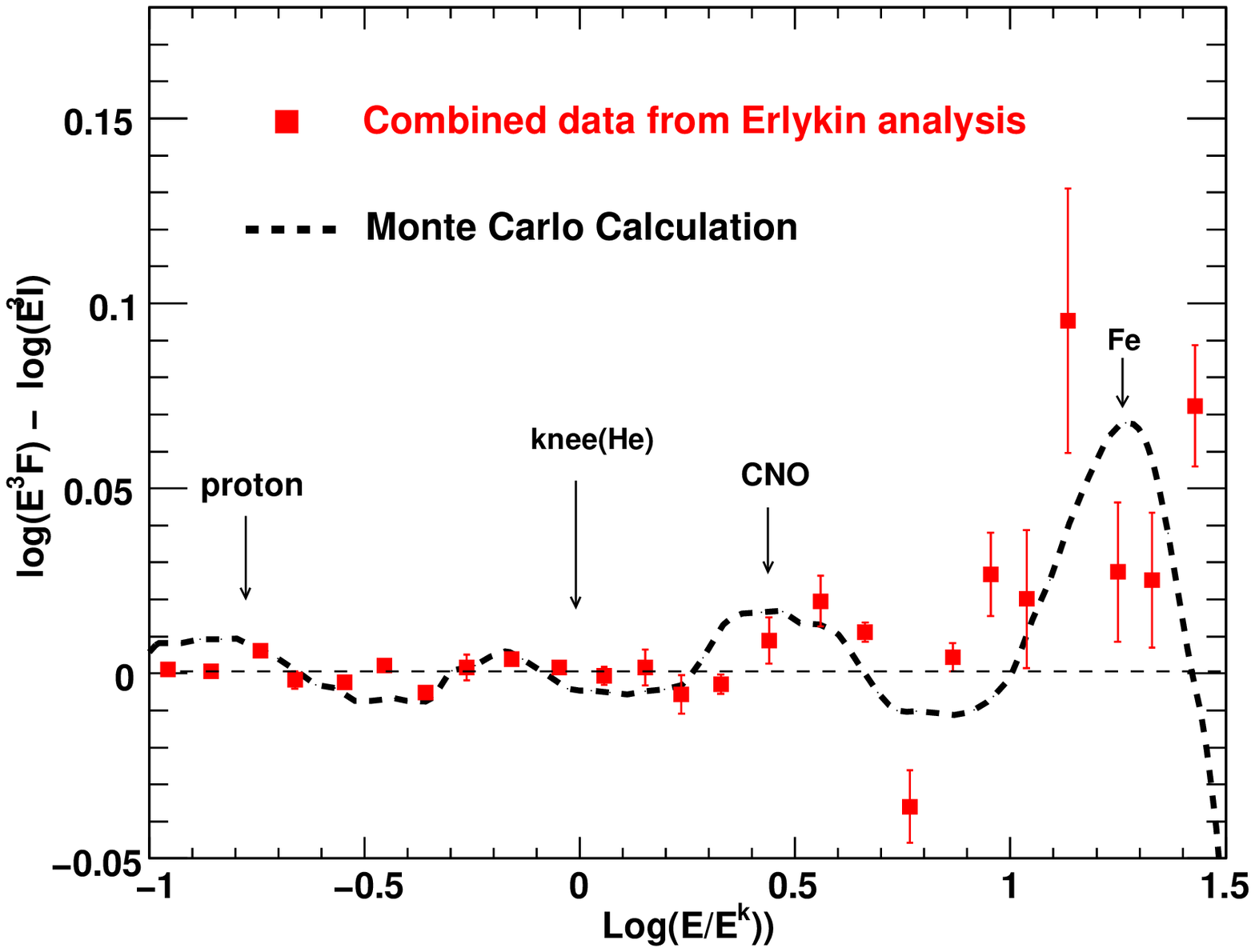}
\caption{\small The fine structures of CR spectrum: theory versus
data. $F$ means the observational or calculated flux of the total CR
spectra, while $I$ indicates the fitting to $F$ according to Eq.
(2). The data are combined results of 6 experiments according to
\cite{2009arXiv0906.3949E}. The positions $\log(E/E^{k})\sim -0.7$
and $\log(E/E^{k})\sim 0.0$ correspond to proton and He, and the two
bumps at $\log(E/E^{k})\sim 0.5$ and $\log(E/E^{k})\sim 1.3$
correspond to CNO group and Fe respectively. }
\label{irregularity}
\end{figure}

Fig. \ref{irregularity} shows a more detailed comparison between
theory and data, after subtracting the main behavior as fitted using
Eq.(\ref{DoublePowLaw}), which enables us to explore the fine
structures of the CR spectra. The data points are combined using $6$
experiments, Tibet Air Shower array\citep{2008ApJ...678.1165A},
KASKADE\citep{2009APh....31...86A},
ARAGATS-GAMMA\citep{2008JPhG...35k5201G},
Yakutsk\citep{2009NJPh...11f5008I},MAKET-ANI\citep{2007APh....28...58C}
and TUNKA [56] according to \cite{2009arXiv0906.3949E}. To avoid the
uncertainty of energy calibration among various experiments, we fit
the break energy $E^k$ in Eq.(\ref{DoublePowLaw}) for each data set
and then normalize it to the Tibet result. It is shown in Fig.2 that
the fine structures of the theoretical calculation is well
consistent with the observational data. The position
$\log(E/E^{k})\sim 0.0$ corresponds to He, and the two bumps at
$\log(E/E^{k})\sim 0.5$ and $\log(E/E^{k})\sim 1.3$ should
correspond to CNO group and Fe respectively. The constraints on the
model parameters are not very strong using the present observational
data, e.g. the final results do not change significantly by varying
the temperature parameters by several tens percent.

Finally we check the spectra of $e^{+}e^{-}$ and position fraction
of these new parameters and find almost identical results as that
derived in Paper I. This is reasonable because the basic parameters
of proton and He, which produce the main part of the
electrons/positrons, are almost unchanged.

\section{Conclusion}

In summary we use the pair production interaction model between CR
nuclei and ambient radiation field proposed in Paper I to explain
the features of the CR spectra, including the sharp knee and fine
structures. Results show that the spectra of CRs agree well with the
observations. In our model, the He composition dominates around the
knee at $\sim 4$ PeV. The sharp knee observed by Tibet air shower
array and confirmed by more and more experiments, can be well
reproduced through the pile-up of He particles. In addition, this
model can explain the fine structures of CR spectrum through the
pile-up effects of CNO group and Fe nuclei. As an additional result,
our model can provide a natural explanation of the ``component B''
problem of individual composition in tens of PeV energy range as
suggested in \cite{2005JPhG...31R..95H}. As in Paper I, the
electron/positron excesses can also be explained.

Furthermore we would like to discuss some implications of the model
parameters. We note that only one single set of parameters is enough
to explain the data. As discussed in Paper I and
\cite{2009arXiv0911.3034H}, it may indicate that the sources are
``standard''  which  have similar parameters such as the temperature
evolution of the radiation field, the relative abundances and
spectral indices for individual elements, or the observed fine
structures of CRs spectra are mainly due to one single source,
either nearby or not so nearby but intensive (e.g. possibly the
Galactic center). This work can be regarded as one progress
approaching the origin of CRs.

\acknowledgments We thank A. D. Erlykin, A. W. Wolfendale, Y. H.
Tan, and L. K. Ding for helpful discussions and comments on the
paper. Hong-Bo Hu thanks Wei Wang for long-term help with the
research work. This work is supported by the Ministry of Science and
Technology of China, Natural Sciences Foundation of China (Nos.
10725524 and 10773011), and by the Chinese Academy of Sciences (Nos.
KJCX2-YW-N13, KJCX3-SYW-N2, GJHZ1004).

\bibliographystyle{aa}
\bibliography{refs}
\end{document}